\documentclass[lettersize,journal]{IEEEtran}

\usepackage{verbatim} 
\usepackage{graphicx}
\usepackage{booktabs}
\usepackage{pgf-pie}
\usepackage{hyperref}
\usepackage{tikz}
\usetikzlibrary{positioning, shapes.multipart, fit, arrows.meta}
\usepackage{subcaption}
\usepackage{multirow}
\usepackage{url}
\usepackage{orcidlink}
\usepackage{xurl}
\usepackage{mdframed}
\usepackage{listings}
\usepackage{xcolor}
\usepackage{algorithm}
\usepackage{amsmath}
\usepackage{bm}
\usepackage{algorithm} 
\usepackage[noend]{algpseudocode} 
\usepackage{xspace}
\usepackage{listings, xcolor}
\usepackage{enumitem}
\usepackage{microtype}
\usepackage{booktabs}
\usepackage{makecell}
\usepackage{varwidth}
\usepackage{amssymb}

\usepackage{siunitx}
\usepackage[capitalize]{cleveref}
\crefname{section}{Sect.}{Sects.}
\Crefname{section}{Section}{Sections}
\crefname{definition}{Def.}{Defs.}
\Crefname{definition}{Definition}{Definitions}
\crefname{algorithm}{Alg.}{Algs.}
\Crefname{algorithm}{Algorithm}{Algorithms}
\lstdefinelanguage{Solidity}{
  keywords={function, external, payable, emit},
  keywordstyle=\color{blue}\bfseries,
  ndkeywords={},
  ndkeywordstyle=\color{darkgray}\bfseries,
  identifierstyle=\color{black},
  sensitive=false,
  comment=[l]{//},
  morecomment=[s]{/*}{*/},
  commentstyle=\color{purple}\ttfamily,
  stringstyle=\color{red}\ttfamily,
  morestring=[b]',
  morestring=[b]"
}

\renewcommand{\paragraph}[1]{\vskip 0.05in \noindent\textbf{#1.}}

\ifCLASSOPTIONcompsoc

  \usepackage[nocompress]{cite}
\else
  \usepackage{cite}
\fi

\ifCLASSINFOpdf

\else

\fi

\newcommand{\tool}[0]{\textsc{PEventCatcher}\xspace}

\begin{document}

\title{Phantom Events: Demystifying the Issues of Log Forgery in Blockchain}

\author{
    Yixuan Liu$^{\orcidlink{0009-0006-2255-7901}}$, 
    Yuxin Dong$^{\orcidlink{0009-0003-6239-4690}}$,
    Ye Liu$^{\orcidlink{0000-0001-6709-3721}}$,
    Xiapu Luo$^{\orcidlink{0000-0002-9082-3208}}$,~\IEEEmembership{Senior Member,~IEEE,}
    and Yi Li$^{\orcidlink{0000-0003-4562-8208}}$,~\IEEEmembership{Member,~IEEE} \thanks{Yixuan Liu and Yi Li are with College of Computing and Data Science, Nanyang Technological University, Singapore.}
    \thanks{Yuxin Dong is with School of Software and Microelectronics, Peking University, Beijing, China.}
    \thanks{Ye Liu is with School of Computing and Information Systems, Singapore Management University, Singapore.}
    \thanks{Xiapu Luo is with the Department of Computing, The Hong Kong Polytechnic University, Hong Kong.}
    
}

\maketitle

\begin{abstract}
With the rapid development of blockchain technology, transaction logs play a central role in
various applications, including decentralized exchanges, wallets, cross-chain bridges, and other
third-party services. However, these logs, particularly those based on smart contract events, are
highly susceptible to manipulation and forgery, creating substantial security risks across the
ecosystem. To address this issue, we present the first in-depth security analysis of transaction
log forgery in EVM-based blockchains, a phenomenon we term Phantom Events. We systematically model
five types of attacks and propose a tool designed to detect event forgery vulnerabilities in smart
contracts. Our evaluation demonstrates that our approach outperforms existing tools in identifying
potential phantom events. Furthermore, we have successfully identified real-world instances for all
five types of attacks across multiple decentralized applications. Finally, we call on community
developers to take proactive steps to address these critical security vulnerabilities.
\end{abstract}

\IEEEpeerreviewmaketitle

\section{Introduction}
\label{Sec:Introduction}

Blockchain is a decentralized digital ledger technology that securely records and verifies transactions across multiple computers, ensuring data integrity and transparency~\cite{bitcoin}. Bitcoin (BTC), the most prominent token in the blockchain ecosystem, reached a remarkable market capitalization of \$1,800 billion in November 2024, demonstrating the impact and success of this technology~\cite{coinmarketcap}.

Building on blockchain technology, Decentralized Applications (DApps) have emerged as a key innovation~\cite{buterin2014next}. These applications run on smart contracts deployed across various blockchain networks, such as Ethereum, Polygon, and Binance Smart Chain (BSC). The expansion of DApps reflects a shift toward decentralized application development, providing diverse services such as Decentralized Finance (DeFi) and Gamefi. This evolving ecosystem is supported by a wide infrastructure, including cryptocurrency wallets and blockchain explorers, which contribute to the dynamism of the blockchain landscape~\cite{tang2022blockchain}.

Smart contract events and transaction logs are essential for tracking actions within blockchain applications, supporting functionalities in cross-chain bridges, wallets, and NFT marketplaces. Falsifying these logs can undermine security, leading to asset theft, fraud, and interoperability issues that threaten user trust. While past research has examined vulnerabilities in specific domains, such as cross-chain bridges~\cite{Xscope,liao2024smartaxe,wang2024xguard}, NFT ownership~\cite{sleepminting}, code inconsistencies~\cite{doccon}, and token behaviors~\cite{TokenScope}, a comprehensive study on event forgery across multiple applications remains absent. This paper addresses this gap by introducing \emph{Phantom Events}, a new class of vulnerabilities that reveal the widespread risks of \emph{log forgery} on the blockchain. Unlike previous studies on isolated scenarios, our work systematically summarizes event misuse across decentralized applications and explores distinct detection methods to identify phantom events. In prior studies, detection tools such as XScope~\cite{Xscope} and XGuard~\cite{wang2024xguard} operate at the source code level, limiting detection when only bytecode is available. Conversely, SmartAxe~\cite{liao2024smartaxe} and TokenScope~\cite{TokenScope} analyze bytecode but their bytecode-only approach misses event parameter details. Guidi et al.~\cite{sleepminting}, which focuses solely on transaction logs, experiences high false-positive rates due to limited contextual analysis. Our tool combines bytecode, source code (when available), and transaction data, enabling comprehensive detection across multiple vulnerabilities while reducing the false positives that affect single-layer analysis methods.

Phantom events are subtle manipulations within blockchain systems that turn normal contract events
into hidden channels for unauthorized actions. These events compromise trust in blockchain
transactions, enabling unauthorized activities that bypass typical security checks. They either
mimic or slightly alter smart contract events to mislead event listeners, user interfaces, and
blockchain analysis tools. This exploitation damages data integrity and can lead to substantial
financial and reputational losses.
For example, on 9 February 2024, attackers caused a 1.04 million USD loss by forging transfer
events~\cite{zero_transfer}.

Our research focuses on understanding and categorizing the various types of phantom events.
We conducted a comprehensive survey of recent blockchain security reports and articles to gather
real-world attacks that exploit vulnerabilities related to phantom events.
Furthermore, we audited multiple active smart contracts and developed a security model to process
blockchain events (\cref{sec:threat}), which serves as the foundation for analyzing and
categorizing the collected attack scenarios (\cref{sec:vector}).
Furthermore, we developed a tool named \tool, designed to detect vulnerabilities related to phantom
events (\cref{sec:detection}).
Designing \tool presented several challenges: (1) many contracts exist only as bytecode, lacking
event semantics, which complicates analysis; (2) phantom events mimic the format of legitimate
events, making them hard to discern; and (3) understanding the business logic of contracts is
necessary to determine the presence of irregularities.

Using our security model and detection tool, we identified numerous vulnerabilities and potential
risks in various blockchain applications (details reported in \cref{sec:evaluation}).
Our results reveal that phantom event vulnerabilities are widespread and affect a variety of
applications in the blockchain ecosystem. Through our auditing efforts, we discovered previously
unreported instances of inconsistent logging vulnerabilities, where on-chain issues resulted in
discrepancies with off-chain records.
Furthermore, our tool identified historical event issues on chain that had gone undetected,
underscoring the effectiveness of the tool in uncovering past incidents.
In contract-level analysis, our tool outperformed existing solutions, demonstrating superior
accuracy in detecting phantom event vulnerabilities.
Furthermore, we identified security issues in multiple real-world applications, including
cryptocurrency wallets, blockchain explorers, and cross-chain bridges, and reported these findings
to project teams, claiming bounties for responsible disclosure.
Finally, we propose mitigation strategies to address these vulnerabilities in~\cref{sec:mitigation}.

In short, we make the following contributions in this paper:
\begin{itemize}
\item \textbf{Novel Attack Taxonomy.}
We are among the first to systematically analyze and categorize attacks
related to the forgery of blockchain transaction logs. We propose a security model that captures
five types of attack. We also identified previously unreported inconsistent logging
vulnerabilities, where on-chain issues result in discrepancies with off-chain records.
\item \textbf{Efficient Detection.}
We developed a detection mechanism that surpasses existing tools in identifying Phantom Events,
successfully uncovering previously unreported attack transactions.
\item \textbf{Practical Implications.}
We identified real-world issues across various blockchain applications, including blockchain
explorers, cryptocurrency wallets, GameFi, DeFi, and NFT marketplaces.
Notably, we found a vulnerability in a cross-chain relayer, which is also used by other projects,
including one with a market capitalization of over \$250 million.
To date, we have reported six cryptocurrency wallet issues, one blockchain explorer issue, and one
cross-chain bridge issue to project teams and bug bounty platforms, with five confirmed and one
resolved, earning a total of \$600 bounty for responsible disclosure.
\end{itemize}

\section{Background}\label{sec:Background}

\subsection{Smart contract event and transaction log in EVM-based Blockchains}
Smart contracts, written in high-level languages like Solidity~\cite{Solidity} and Vyper~\cite{Vyper}, are compiled into bytecode for execution on the Ethereum Virtual Machine (EVM)~\cite{EVM}. EVM-based blockchains support a wide range of functionalities, from simple transfers to complex DApps~\cite{DApp}. In these blockchains, \textit{events} and \textit{transaction logs} enable communication between smart contracts and external systems. When a smart contract emits an event, it is recorded as a transaction log, involving key components as follows.

\paragraph{Event}
An event, such as \texttt{Deposit}, is defined as a tuple $E_{\texttt{Deposit}}(P_1, P_2, \ldots, P_n)$, where the parameters $P_i$ may be \textit{indexed} (stored in \textit{topics}) or \textit{non-indexed} (stored in \textit{data}). For example, in the event
\texttt{Deposit(address indexed account, uint256 amount, uint256 timestamp);},
the \texttt{account} is an \textit{indexed} parameter, which means it will be included in the event's topics, enabling more efficient searching and filtering in the event logs. The \texttt{amount} and \texttt{timestamp} parameters are not indexed, but they still provide valuable information about the transaction.


\paragraph{Listener}
An external system (e.g., a DApp or wallet) that monitors specific events by subscribing to logs through RPC methods.

\paragraph{Transaction}
A transaction $TX_{\textit{hash}}$ represents an action on the blockchain, such as invoking a function $f_{\texttt{name}}$ within a smart contract. It includes the sender’s and recipient’s addresses, the transfer amount, and input data defining the function call and its parameters.

\paragraph{Sender}
The sender \( TX_{\textit{sender}} \) is the user address that creates the transaction.

\paragraph{Emitter}
The emitter $S_{\texttt{address}}$ is the smart contract triggering the event upon meeting certain conditions.

\paragraph{Transaction Log}
A log \( L_{hash} = (Topics_0^i, \allowbreak{} Data^i,\allowbreak{} S_{\texttt{address}}^i) \) is
generated
when an event is emitted during smart contract execution. \( Topics_0^i \) represents the event
signature for the \(i\)-th log, \( Data^i \) contains non-indexed parameters, and \(
S_{\texttt{address}}^i \) is the contract emitting the event. These logs are stored on the
blockchain and referenced for event monitoring.


\subsection{Token}
In EVM-based blockchains, tokens are implemented as smart contracts adhering to standards like \textit{ERC-20} and \textit{ERC-721}, which define interfaces and events to ensure compatibility across DApps. For instance, ERC-20 specifies events like \texttt{Transfer} and \texttt{Approval} for tracking token transfers and approvals, while \textit{ERC-721} defines similar events but is adapted for NFTs, where each token is unique. These standardized events enable external systems, such as wallets and exchanges, to monitor token activities and update user balances or ownership records in real time, ensuring accurate and efficient tracking.

\begin{figure}[t]
    \centering
    \includegraphics[width=\linewidth]{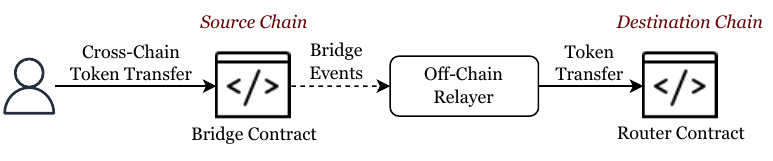}
    \caption{The basic architecture of a cross-chain bridge.}
    \label{fig:Bridge}
\end{figure}

\subsection{Cross-Chain Brdiges}
Some DApps rely on off-chain modules to monitor transactions and events emitted by smart contracts in order to stay updated on on-chain activities and state changes in real time. For example, cross-chain bridges use these events to trigger off-chain processes, update user interfaces, or others.

In the case of a cross-chain bridge, as illustrated in \cref{fig:Bridge}, the smart contract on \emph{source chain} acts as the emitter $S_{source}$, emitting specific events such as \texttt{Deposit} when users deposit assets. And the smart contract on the \emph{destination chain} as an actor, being invoked by an off-chain relayer to process the deposited assets and complete the transfer operation.

When a user deposits tokens in a \emph{bridge contract} in the source chain, the deposit event $E_{\texttt{Deposit}}$ is emitted and recorded as a transaction log $L_{deposit_{tx}}$. An \emph{off-chain relayer}, functioning as a listener, monitors these logs to detect the \texttt{Deposit} event, process the event information, and coordinate with the smart contract on the destination chain to complete the asset transfer.
\section{Threat Model and Motivating Examples}\label{sec:threat}

This section presents a threat model for the systematic analysis of event-related attacks and illustrates the impact of Phantom Events through motivating examples.
\subsection{Trust Boundaries in Event Interactions}

In the complex environment of blockchain interactions, establishing clear trust boundaries is essential for comprehensively understanding the security risks and vulnerabilities associated with transaction and event workflows. The interact model illustrated in \cref{fig:system model} defines three primary trust boundaries:
\begin{figure}[t]
    \centering
    \includegraphics[width=.9\linewidth]{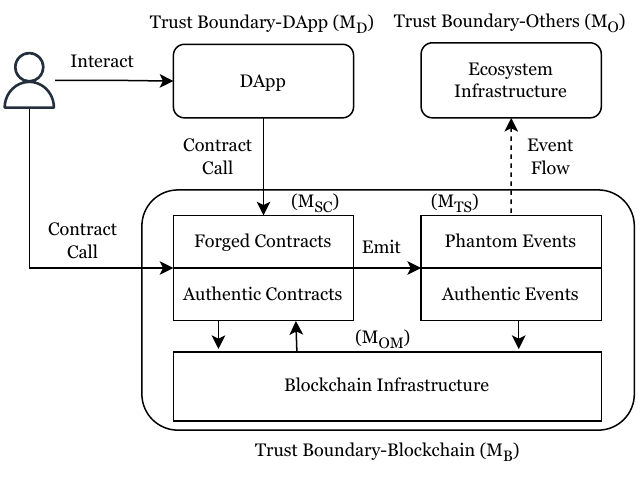}
    \caption{Trust Boundaries in Blockchain Interactions}
    \label{fig:system model}
\end{figure}

\paragraph{Trust Boundary-Blockchain (\(\bm{\mathcal{M}}_B\))}
The core trust boundary is the blockchain, which serves as the backbone of the system where transactions and smart contract events are stored.
Within this boundary lies the Transaction Storage (\(\bm{\mathcal{M}}_{TS}\)), which can be divided into \emph{phantom events} and \emph{authentic events}.
Phantom events are artificially created or manipulated events that may trigger unintended actions, while authentic events are the expected output of blockchain transactions.
The Smart Contract area (\(\bm{\mathcal{M}}_{SC}\)) can be divided into \emph{authentic contracts} and \emph{forged contracts}, indicating the potential for contracts to operate as intended or act maliciously.
Other Blockchain Modules (\(\bm{\mathcal{M}}_{OM}\)) include consensus mechanisms, node communication protocols, and other components that support blockchain functionality.

\paragraph{Trust Boundary-DApp (\(\bm{\mathcal{M}}_{D}\))}
The second boundary encompasses DApps that act as interfaces between users and the blockchain.
DApps listen for events from the blockchain and respond accordingly, often triggering transactions or updates in response to these events.

\paragraph{Trust Boundary-Others (\(\bm{\mathcal{M}}_{O}\))}
The third boundary includes the broader ecosystem, such as off-chain services, external APIs, and wallets. These components also rely on the integrity of blockchain events for proper functionality.

\subsection{Threat Model}
In our threat model, attackers seek to compromise DApps and the broader ecosystem infrastructure by emitting forged events, and may also exploit these attacks to mislead users through social engineering tactics. While regular users emit legitimate events by invoking authentic contracts, attackers can emit forged events in several ways: by calling a forged contract, by calling an authentic contract, or by invoking a forged contract that subsequently calls an authentic contract to emit a forged event. The attacker's capabilities are confined to making direct contract calls or interacting with DApps.

\subsection{Motivating Examples}\label{sec:motivation}
Numerous real-world blockchain attacks have occurred due to the forgery of transaction logs.
For example, the Qubit Bridge and pNetwork Bridge attacks~\cite{Qubit,pNetwork} caused \$80 million and \$4.3 million losses, respectively, due to the exploitation of Phantom Event vulnerabilities.

\begin{figure}[h]
    \centering
    \includegraphics[width=1\columnwidth]{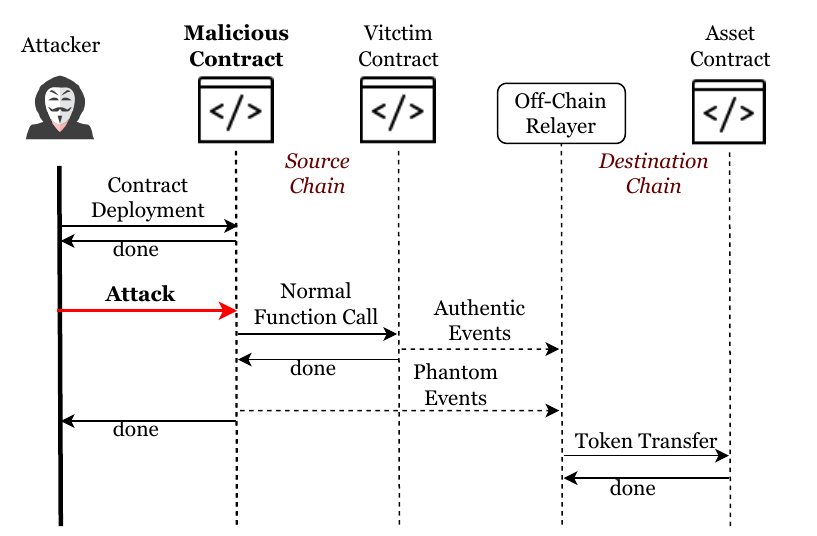}
    \caption{Attack sequence of pNetwork Hack.}
    \label{fig:sqeuence}
\end{figure}

In the pNetwork hack, there is a flaw in the event handling mechanism, where if both a malicious
contract and a legitimate contract are invoked within the same transaction and both emit events,
the mechanism mistakenly treats all emitted events as legitimate contract events. So as shown in
the attack sequence (\cref{fig:sqeuence}), the attacker first deployed a malicious contract
$S_{\texttt{malicious}}$ on the source chain which invoking the legitimate contract
$S_{\texttt{legitimate}}$ through a function call. After the legitimate contract
$S_{\texttt{legitimate}}$ executed its function $f_{\texttt{deposit}}$ and emitted the authentic
event $E_{\texttt{Deposit}}$, the malicious contract $S_{\texttt{malicious}}$ simultaneously
emitted a counterfeit event $E_{\texttt{Deposit}}$ with an untruthful amount.
Since both the authentic event and the counterfeit event were recorded under the same transaction
$TX_{\textit{deposit}}$, the relayer processed the counterfeit event as if it were legitimate.
This allowed the attacker to transfer unauthorized funds to the destination chain by exploiting the
inability of the system to differentiate between the two events.

In the Qubit Bridge attack, the attacker exploits a flaw in the \texttt{deposit} function of the
victim contract, as illustrated in \cref{code:1}.
The relayer transfers assets on the destination chain by retrieving the token address and amount
from the event emitted on the source chain. However, the attacker exploited this by using the
\texttt{deposit} function to emit an event that should only be emitted by the \texttt{depositETH}
function, allowing them to trigger a transfer of ETH on the destination chain without actually
depositing any ETH on the source chain, thus bypassing the relayer's checks and profiting from the
attack.

\paragraph{Idea of \tool}
Motivated by these real-world examples, we apply customized detection methods for various types of
vulnerabilities.
The phantom events may arise from issues in smart contract logic or from flaws in off-chain
processors.
To address this, we designed a smart contract-level approach, which identifies vulnerabilities that
could lead to phantom events.
Additionally, we developed an on-chain approach for transaction data, inspecting transactions for
patterns or anomalies that may indicate attacks.
These approaches complement each other and enable a comprehensive detection framework that
addresses both contract-based and transaction-based origins of phantom events.
\section{Attack Taxonomy and Analysis}\label{sec:vector}

In this section, we present our taxonomy of attacks caused by phantom events and provide detailed
explanations as well as detection rules for each attack.

\begin{table*}[ht]
\centering
\caption{Mapping relationship between attack vectors and related modules.}
\label{tab:attack-classification}
\begin{tabular}{l|l|p{8.5cm}|p{3cm}}
\toprule
\textbf{Level} & \textbf{Attacks} & \textbf{Descriptions} & \textbf{Related Modules} \\
\midrule
\multirow{2}{*}{\textbf{On-chain}}
& event counterfeiting & Use existing contracts to emit Phantom events & \(\bm{\mathcal{M}}_{TS}\), \(\bm{\mathcal{M}}_{O}\) \\
\cmidrule{2-4}
& inconsistent logging & Inconsistencies between database and blockchain event emissions & \(\bm{\mathcal{M}}_{SC}\), \(\bm{\mathcal{M}}_{TS}\), \(\bm{\mathcal{M}}_{D}\) \\
\midrule
\multirow{3}{*}{\textbf{Off-chain}}
&contract imitation & Deploy a malicious contract to emit Phantom events & \(\bm{\mathcal{M}}_{SC}\), \(\bm{\mathcal{M}}_{TS}\), \(\bm{\mathcal{M}}_{O}\) \\
\cmidrule{2-4}
& transfer event spoofing & Emit Phantom events with social engineering attack & \(\bm{\mathcal{M}}_{SC}\), \(\bm{\mathcal{M}}_{TS}\), \(\bm{\mathcal{M}}_{O}\) \\
\cmidrule{2-4}
& event handling error & Incorrect display/ insertion/storage due to faulty event processing & \(\bm{\mathcal{M}}_{D}\), \(\bm{\mathcal{M}}_{O}\) \\
\bottomrule
\end{tabular}
\end{table*}

We relied on industry reports, academic literature, and real-world smart contract audits to construct a taxonomy of relatively new and understudied phantom event attacks. We model five distinct attack
scenarios, which are classified into two categories based on their origins:
(1) \emph{on-chain (smart contract) vulnerabilities} and (2) \emph{off-chain weaknesses}.
These attacks are summarized in our attack classification table (\cref{tab:attack-classification}).

Detailed demonstrations of these attacks are available in the supplementary materials.\footnote{\url{https://github.com/PhantomEvent/Event-attack-demo}}

\subsection{Attack Vectors and Detection Rules}
This section introduces five attacks related to phantom events and proposes corresponding detection rules. Examples of these attacks can be found in~\cref{appendix:examples}.

\subsubsection{Event Counterfeiting}\label{vec:attack1}

\begin{figure}
\small
\footnotesize
\begin{lstlisting}[language=Solidity]
function depositETH(uint destinationChainId) external payable {
  require(msg.value > 0, "Deposit amount must be greater than 0");
  ethBalances[msg.sender] += msg.value;
  emit Deposit(msg.sender, msg.value, address(0), destinationChainId);
}

function deposit(address token, uint amount, uint destinationChainId) external {
  require(amount > 0, "Deposit amount must be greater than 0");
  safeTransfer(token, address(this), amount);
  tokenBalances[msg.sender][token] += amount;
  emit Deposit(msg.sender, amount, token, destinationChainId);
}

function safeTransfer(address token, address to, uint value) internal {
  (bool success, bytes memory data) = token.call(abi.encodeWithSelector(0xa9059cbb, to, value));
  require(success && (data.length == 0 || abi.decode(data, (bool))), "!safeTransfer");
}
\end{lstlisting}
\caption{Proof of concept for the event counterfeiting attack, where \texttt{deposit} and
\texttt{depositETH} could emit the same event.}\label{code:1}
\end{figure}

This attack takes advantage of legitimate contracts that emit the same event from multiple functions or execution paths (\(\bm{\mathcal{M}}_{TS}\)). In the provided code example~\cref{code:1}, both the \texttt{depositETH} and \texttt{deposit} functions emit a \texttt{Deposit} event, where the third parameter signifies the token type. While \texttt{depositETH} uses \texttt{address(0)} to represent ETH, the \texttt{deposit} function uses the token address for ERC-20 tokens. However, the \texttt{deposit} function can also emit an event with \texttt{address(0)}, mimicking an ETH deposit.

An attacker could exploit this by calling the \texttt{deposit} function and forging the third parameter as \texttt{address(0)}, without actually transferring any ETH. This creates a \emph{phantom} \texttt{Deposit} event that resembles a real ETH deposit. Off-chain validation systems, which rely on event logs for transaction verification, may mistakenly treat this event as a legitimate ETH deposit, allowing the attacker to fraudulently claim large sums on the destination chain due to the flawed event validation process.

\paragraph{Detection Method}
This vulnerability occurs when different functions or execution paths emit the same event with identical parameters, even though each function imposes different constraints. Off-chain systems often treat these events as authentic without verifying the parameter constraints. If a function emits a value that is expected to be constrained or validated by another function, it is considered a potential vulnerability. The detection rule checks for cases where identical parameter values are emitted along different execution paths, even when those paths should enforce distinct constraints, and flags such events as potential vulnerabilities.

The formalized detection rule is expressed as follows.

\begin{equation}
\small
\text{EC} = \left\{
e \in E \mid
\begin{aligned}
  & V(f, e) \cap V(g, e) \neq \emptyset \text{ and } \\
  & \exists \; f, g \in F(e)
\end{aligned}
\right\}
\end{equation}

\begin{itemize}
    \item \( E \) is the set of emitted events, and \( e \in E \) represents an individual event.
    \item \( F \) is the set of all function paths in the contract that can emit events, and \( f, g \in F \) are specific function paths that emit the same event \( e \).
    \item \( V(f, e) \) and \( V(g, e) \) represent the range of values that the parameters of event \( e \) can take along function paths \( f \) and \( g \).
\end{itemize}

\subsubsection{Inconsistent Logging}\label{vec:attack2}

\begin{figure}
\footnotesize
\small
\begin{lstlisting}[language=Solidity]
function requestWithdraw(uint256 _type, uint256 _amount) external {
  require(WITHDRAW_ALLOWED, "TroyEmpire: Withdrawal is disabled for now");
  emit WithdrawalRequested(_msgSender(), _type, _amount);
}
\end{lstlisting}
\caption{Proof of concept for the inconsistent logging attack.}\label{code:2}
\end{figure}

Many DApps adopt a hybrid logging model that uses both blockchain and traditional databases to record critical operations. For example, DeFi platforms record financial transactions, such as deposits and withdrawals, while GameFi platforms track user operations. Our analysis revealed that numerous projects not only rely on database records, but also emit logs on the blockchain.

Specifically, poorly designed smart contracts may allow attackers to emit forged events, leading to inconsistencies between the blockchain and the database logs. For example, a financial platform might use both blockchain and off-chain databases to log withdrawal requests. If the smart contract lacks proper access controls, attackers could arbitrarily generate withdrawal events on the blockchain, creating a mismatch with the database records, as shown in \cref{code:2}.

\paragraph{Detection Method}
This vulnerability occurs when smart contracts emit events without proper access control or validation against stored data. Specifically, some contracts lack necessary constraints, allowing any user to trigger critical events such as withdrawals without validating the event parameters against the contract's state. To detect this, we analyze the contract's functions for two key aspects: (1) the presence of access control mechanisms or event parameter validation (e.g., \texttt{require} statements) that restrict who can emit events, and (2) whether the function interacts with storage (i.e., reads from or writes to storage variables) when emitting events to ensure that parameters are validated against previously stored values. If a function emits an event without adequate access control or does not interact with storage, it is flagged as a potential vulnerability.

The formalized detection rule is expressed as follows.

\begin{equation}
\small
\text{IL} = \left\{
f \in F \mid
\begin{aligned}
  & Constraint(f) = \emptyset \text{ or } \\
  & (S_{\texttt{read}}(f) = \emptyset \text{ and } S_{\texttt{write}}(f) = \emptyset)
\end{aligned}
\right\}
\end{equation}

\begin{itemize}
    \item \( Constraint(f) \): The access control mechanism for function \( f \) or constraint for event parameters. If \( Constraint(f) = \emptyset \), there are no access controls or constraints.
    \item \( S_{\texttt{read}}(f) \) and \( S_{\texttt{write}}(f) \): The storage variables read and written by the function \( f \).
\end{itemize}

\subsubsection{Contract Imitation}\label{vec:attack3}

This attack involves the deployment of a malicious contract to exploit or imitate an existing contract (\(\bm{\mathcal{M}}_{SC}\)). We classify this attack into two subtypes based on its characteristics. The first subtype, called Blended Event Attack, occurs when a malicious contract interacts with a legitimate contract, causing events from both contracts to be logged within the same transaction. This blending of events makes it difficult to distinguish between legitimate and fraudulent activity. The second subtype, named Mimicry Contract Attack, involves an attacker deploying a forged contract that imitates the behavior of a legitimate contract, allowing the attacker to manipulate event logs and transaction details.

An \emph{authentic} contract typically records logs from its own functions. However, most contracts permit external calls, allowing malicious contracts to invoke their functions, leading to logs from both contracts being recorded in a single transaction (\(\bm{\mathcal{M}}_{TS}\)). This scenario, termed the \emph{Blended Event} attack, introduces a security risk when the log emitter is not verified.

The \emph{Mimicry Contract} attack exploits transparency practices, as many DApp teams publish their contract source code to promote trust. This allows attackers to modify and redeploy code, creating malicious contracts that mimic legitimate ones and emit event logs that can be freely manipulated (\(\bm{\mathcal{M}}_{TS}\)). For ERC-20 tokens or NFTs, this manipulation enables attackers to forge transaction records, such as minting or transferring tokens or NFTs, which can include falsifying the sender to make an NFT appear as though it was issued by a well-known artist.

\paragraph{Detection Method}
To detect this, we analyze the \emph{transaction logs}. Specifically, we check whether the same event signature (\( Topics_0 \)) appears multiple times in the same transaction, but is emitted by different contracts. If the event signature is emitted by both the authentic contract and a forged contract within the same transaction, it is flagged as a potential attack transaction. This detection rule helps identify cases where the event signatures match but are emitted by different emitters, indicating a \emph{Blended Event Attack}.

\begin{equation}
\small
\text{BE} = \left\{
L_{\textit{hash}} \mid
\begin{aligned}
  & Topics_0^i = Topics_0^j \text{ and } \\
  & S_{\texttt{address}}^i = S_{\texttt{auth}} \text{ and } \\
  & S_{\texttt{address}}^j = S_{\texttt{forge}}
\end{aligned}
\right\}
\end{equation}

\subsubsection{Transfer Event Spoofing}\label{vec:attack4}

This attack is a form of social engineering attack~\cite{social_engineer}. One common example is the Zero Transfer Scam, where the attacker creates a phantom event that mimics a legitimate token transfer. After the user sends tokens, the attacker generates a fake event that makes it appear as though the user has sent tokens to a similarly named recipient address, which belongs to the attacker. This counterfeit event is recorded by wallets and blockchain explorers, misleading the victim into transferring additional funds to the attacker's account. This scam has reportedly caused losses of at least \$27.36 million USD, affecting 28,414 victims~\cite{wallet_visual}.

The second variation, called the Airdrop Scam~\cite{spoof}, exploits the common Web3 marketing tactic of token airdrops. Attackers forge phantom events by spoofing the sender’s address using real ERC-20 or ERC-721 contracts, often selecting addresses designed to catch attention (e.g., 0x8888...8888). These phantom airdrops trick victims into interacting with the tokens. Attackers may set up honeypot contracts where users can buy tokens but are unable to sell them, or use phantom event logs to promote malicious links (e.g., using ENS names to deceive victims). In more severe cases, attackers may perform rug pulls, withdrawing liquidity and leaving victims with worthless tokens~\cite{rugpull}.

\paragraph{Detection Method}
Attackers exploit the fact that many third-party services, such as wallets and blockchain explorers, blindly trust events emitted by smart contracts without verifying their authenticity. By taking advantage of this lack of verification, attackers can generate phantom events that mimic legitimate transfers, misleading users into transferring funds to the attacker's address. To detect this, we can analyze the transaction logs to ensure that the transfer event was genuinely initiated by the correct sender, or verify if the sender had approved the address that initiated the transaction.

\begin{equation}
\small
\text{TS} = \left\{
L_{\textit{hash}} \mid
\begin{aligned}
  & TX_{\textit{sender}} \neq \textit{TokenSender} \text{ and } \\
  & Approve(\textit{TokenSender}, TX_{\textit{sender}}) = \text{false}
\end{aligned}
\right\}
\end{equation}

\begin{itemize}
    \item \( \textit{TokenSender} \) represents the sender of the tokens in event.
    \item \( Approve(\textit{TokenSender}, TX_{\textit{sender}}) \) indicates whether the token sender has authorized the transaction initiator to transfer their tokens.
\end{itemize}

\subsubsection{Event Handling Error}\label{vec:attack5}
This attack exploits vulnerabilities in how blockchain explorers, wallets, and monitoring tools process and interpret transaction data. These applications rely on on-chain data to provide real-time information to users. When phantom events are mistakenly treated as legitimate, this can result in inaccurate wallet balances, false transaction histories, or incorrect asset ownership records, which misleads users and potentially impacts their decision-making.

For example, in the context of \emph{Contract Imitation} attacks, even if phantom events are not specifically targeting blockchain explorers, they may still cause explorers to log these events as valid ERC-20 or NFT transfers, despite no actual transfer occurring. This misinterpretation allows attackers to create the appearance of transfers without underlying transactions.

Attackers can further exploit this by embedding malicious payloads within event data to launch attacks such as cross-site scripting (XSS) or SQL injection (SQLi). Without proper sanitization, such payloads can lead to unauthorized actions on the DApp interface, data theft, or database compromise.

\paragraph{Detection Method}
This attack leverages the implicit trust that blockchain explorers, wallets, and monitoring tools place in on-chain data. Effective detection requires adaptable methods tailored to the needs of each off-chain application. Explorers should validate events against actual token transfers, while wallets need to confirm that transfers are permitted for the sender. By customizing detection rules based on these unique validation standards, applications can mitigate the risks posed by phantom events and embedded malicious payloads.

\section{Vulnerability Detection}\label{sec:detection}

\begin{figure*}[t]
  \centering
  \includegraphics[width=\linewidth]{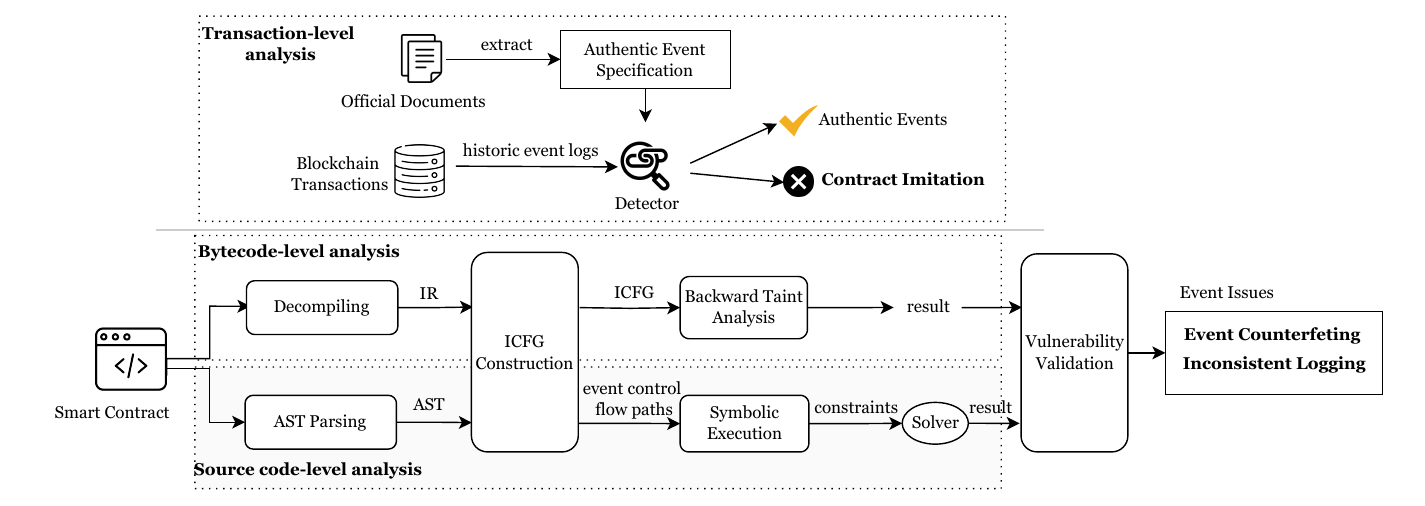}
  \caption{The architecture of \tool.}
  \label{fig:tool}
\end{figure*}

In this section, we focus on detecting vulnerabilities related to the attack vectors discussed in
\cref{sec:vector}. We introduce our hybrid approach, which combines static analysis with symbolic
execution and on-chain data monitoring to ensure comprehensive vulnerability detection.

\subsection{\tool Framework Design}

As shown in \cref{fig:tool}, our detection approach employs a multi-level strategy, encompassing
transaction-level, bytecode-level, and source code-level for vulnerability detection.

At the transaction level, we manually researched documentation for different projects to establish appropriate rules, given the variation in authentic events and emitters. These rules are applied to both historical and real-time on-chain transaction data to detect \emph{Contract Imitation} by identifying instances where phantom events are blended with authentic ones.

For smart contracts, our analysis proceeds across bytecode and source code levels. Initially, we decompile smart contract bytecode into an intermediate representation (IR) in three-address form using Gigahorse~\cite{Gigahorse}. From this IR, we construct an inter-contract control flow graph (ICFG) based on basic blocks. This graph enables backward taint analysis, which we use to trace event-related paths and detect vulnerabilities related to \emph{Event Counterfeiting} and \emph{Inconsistent Logging}.

If the source code is available, we extend our analysis using Slither to perform source code-level analysis. This provides a secondary confirmation for \emph{Event Counterfeiting} by tracing event-related call paths and extracting constraints for symbolic execution, allowing us to verify parameter values more comprehensively.

Due to the unique nature of certain attacks, \emph{Event Handling Error} lacks a specific detection method as it heavily relies on off-chain system validation. Similarly, \emph{Transfer Event Spoofing} is primarily a social engineering attack that focuses on token behavior analysis. Therefore, our tool primarily targets detecting the remaining three types of attacks: \emph{Event Counterfeiting}, \emph{Inconsistent Logging}, and \emph{Contract Imitation}, where distinct patterns and contract-level vulnerabilities can be effectively identified and addressed.

\subsection{Transaction-Level Detection}
At the transaction level, off-chain monitoring is applied to analyze on-chain transaction data and detect potential \emph{Contract Imitation} attacks.
By establishing domain-specific rules based on the expected behavior of smart contracts, we systematically verify whether transactions adhere to correct logic and identify any signs of malicious activity that may indicate phantom events or unauthorized interactions.

This analysis involves parsing transaction data and applying a set of rules to validate key aspects of events, including:
(1) ensuring that the emitted event has the correct signature;
(2) verifying that the event emitter matches the expected contract address;
(3) checking that the function execution path aligns with the expected logic; and
(4) confirming that the event parameter values match the expected ones.

By enforcing these rules, we can detect anomalies that may indicate \emph{Contract Imitation} attacks.
If an attack transaction is detected, further analysis is performed to confirm its impact, such as causing unauthorized fund transfers or state changes.

\begin{table*}[t]
  \centering
  \caption{Transaction logs of the pNetwork attack.}
  \label{tab:pnetwork_transaction}
  \resizebox{\textwidth}{!}{
    \begin{tabular}{lllp{7cm}}
      \toprule
      \textbf{Events} & \textbf{Parameters} & \textbf{Emitters} & \textbf{Parameter Values} \\
      \midrule
      Burned &
      operator, from, amount, data, operatorData &
      pBTC &
      (Attacker, Attacker, 0, , \_ ) \\
      \midrule
      Transfer &
      from, to, amount &
      pBTC &
      (Attacker, address(0), 0)\\
      \midrule
      Redeem &
      redeemer, value, underlyingAssetRecipient, userData &
      pBTC &
      (Attacker, 0, Attacker\_Bitcoin\_address, \_) \\
      \midrule
      Redeem &
      redeemer, value, underlyingAssetRecipient, userData &
      Malicious Contract &
      (Attacker, 274[...]144, Attacker\_Bitcoin\_address, \_) \\
      \bottomrule
  \end{tabular}}%
  \end{table*}

For example, in the pNetwork attack~\cref{tab:pnetwork_transaction}, a \emph{Redeem} event was emitted by a malicious contract with an unexpected contract address on the source chain. Applying these transaction-level rules allowed us to detect the attack. Subsequent monitoring of the destination chain transaction confirmed that the attack succeeded with unauthorized fund transfers.

\subsection{Smart-Contract-Level Detection}

\begin{algorithm}[t]
\label{algo:bytecode}
  \footnotesize
  \caption{Detecting Inconsistent Logging and Event Counterfeiting via Backward Taint Analysis}
  \textbf{Input:} \textit{IR}, an intermediate representation of smart contract bytecode via decompilation. \\
  \textbf{Output:} $E_v$, a set of vulnerable events. 
  \begin{algorithmic}[1]
    \State  $E_v = \emptyset$. 
    \State \textbf{construct} \textit{ICFG} out of IR.
    \State \textbf{extract} \textit{LogOps}, i.e., event log operations, from IR. \Comment{Each operation contains a event signature and a set of logging data variables} 
    
    \State  \textit{Vars} $= \emptyset$, 
    \For{logOp $\in$ \textit{LogOps}}
      \State Vars $\gets$ Vars $\cup$ \textbf{vars}(logOp)  \Comment{Record all logging data variables}
      \State Slices $\gets$ \textsc{BackwardSlicing}(logOp, ICFG) \Comment{Each slice is a reverse execution flow path from logOp to function entry point }
          \State \textbf{extract} \textit{Paths}, i.e., inter-functional paths based on Slices and ICFG
      \State taintedPaths $\gets \emptyset$   
      \For{$p \in$ Paths}
        \State $source_{op} \gets$ \textsc{TaintAnalysis}($p$, Vars) 
        \If{$source_{op} \neq null$}
            \If{\textbf{sstore}($p$[:$source_{op}$]) $= \emptyset$} 
                \State $E_v \gets E_v \cup \textbf{event}(logOp)$ \Comment{No storage operations}
            \EndIf
            \State taintedPaths $\gets taintedPaths \cup \{\textbf{event}(logOp)\} $   
        \Else
            \If{HasExternalCall($p$) $\land$ not HasJumpi($p$)}
                 \State $E_v \gets E_v \cup \textbf{event}(logOp)$ \Comment{No constraints}
            \EndIf

        \EndIf
        
      \EndFor
      \If{$||taintedPaths|| >$ 1}
               \State $E_v \gets E_v \cup \textbf{event}(logOp)$ \Comment{A event logging action has multiple tainted paths}
        \EndIf
    \EndFor
    \State \textbf{return} $E_v$

\Procedure{TaintAnalysis}{path, taintVars}  
  \State $entry = path[-1]$ \Comment{Function entry point is the last item of this reverse execution path} \label{begin:taintanalysis}
  \For{$instr \in path$}
    \If{$\textbf{vars}(instr) \cap taintVars \neq \emptyset$}
        \State taintVars $\gets$ taintVars $\cup$ $\textbf{vars}(instr)$  \Comment{Taint propagation}
    \EndIf
  \EndFor

  \If{$vars(entry) \cap taintVars \neq \emptyset $}
    \State \textbf{return} entry
  \Else 
    \State \textbf{return} null  \label{end:taintanalysis}
  \EndIf 
  
\EndProcedure 


  \end{algorithmic}
  \label{algorithm}
\end{algorithm}

At the smart-contract level, we detect vulnerabilities specifically related to \emph{Event Counterfeiting} and \emph{Inconsistent Logging} through a multi-tiered analysis approach consisting of two parts: (1) identifying potential phantom events through \emph{inter-procedural backward taint tracking} at the bytecode level, and (2) confirming vulnerabilities by \emph{validating event parameter constraints} at the source-code level.
This dual-method approach allows us to analyze different representations of the contract to
identify vulnerabilities that may not be visible at a single level of abstraction.

\subsubsection{Phantom Event Identification Through Interprocedural Backward Taint Tracking}
At the bytecode level, as described in~\cref{algorithm}, the detection of vulnerabilities related to \emph{Event Counterfeiting} and \emph{Inconsistent Logging} is performed through backward taint analysis using an intermediate representation (IR) of the smart contract bytecode. The algorithm identifies vulnerable events \(E_v\) by analyzing the flow of data from event log operations to their sources and checking for specific conditions that indicate vulnerability.

The algorithm begins by initializing an empty set \(E_v\) to store vulnerable events (line 1). It constructs an inter-procedural control flow graph (ICFG) from the IR (line 2) and extracts all event log operations (\textit{LogOps}) from the IR (line 3). Each log operation consists of an event signature and a set of event data variables. The algorithm then initializes an empty set \textit{Vars} to record the event variables (line 4).

The construction of the ICFG involves recovering function-level information, such as function names and event signatures, by matching hashed identifiers in the bytecode with entries in a signature database. A control flow graph (CFG) is then constructed using basic blocks that represent possible execution paths within each function. These basic blocks are connected by directed edges based on control flow instructions and jump targets, forming the CFG. To capture interactions across functions, interprocedural calls are identified, and edges are added between caller and callee functions, thereby extending the CFG into an ICFG. 

For each log operation in \textit{LogOps}, the algorithm updates \textit{Vars} with the variables involved in the log operation (line 6). It then performs a \textsc{BackwardSlicing} analysis on the ICFG to extract slices representing the reverse execution flow from the log operation to the function entry point (line 7). These slices are used to derive inter-functional paths (\textit{Paths}) for further analysis (line 8).

For each path \(p\) in \textit{Paths}, the algorithm performs taint analysis using the \textsc{TaintAnalysis} function (line 9). This function evaluates whether tainted data propagates from the log operation to the path entry point. Starting from the log operation, tainted variables are tracked by propagating taint through instructions in the reverse execution path. Specifically, if an instruction interacts with tainted variables, its associated variables are added to the set of tainted variables. The process continues until reaching the path entry point, which serves as the source of the taint. If tainted variables are found at the path entry, the function returns the entry point as the (\(source_{op}\)); otherwise, it returns \texttt{null}.

If a source of tainted data (\(source_{op}\)) is identified, the algorithm checks if there are any storage operations (\texttt{sstore}) in the portion of the path leading to the source (line 11). Additionally, it verifies whether the variables involved in these \texttt{sstore} operations are subject to constraints (e.g., conditional jumps like \texttt{jumpi}). If no \texttt{sstore} operations are found, or if the \texttt{sstore} operations lack constraints on their variables, the corresponding event is added to \(E_v\) as potentially vulnerable to \emph{Inconsistent Logging} (line 12). The event is also recorded in a separate set of tainted paths for further evaluation (line 13).

If no source of tainted data is identified, the algorithm checks if the path contains an external call (\texttt{call}) without corresponding constraints (e.g., conditional branches like \texttt{jumpi}) (line 15). If such conditions are found, the event is added to \(E_v\) as vulnerable to \emph{Event Counterfeiting} (line 16).

After analyzing all paths for a log operation, the algorithm checks if multiple tainted paths exist for the same event (line 18). If so, the event is flagged as susceptible to \emph{Event Counterfeiting} due to the presence of multiple emission paths (line 19).

By analyzing the tainted data flow and identifying conditions such as missing constraints, unchecked external calls, and the absence of storage operations, the algorithm effectively detects vulnerabilities. Events flagged in \(E_v\) are susceptible to either \emph{Event Counterfeiting}, where the same event may be misleadingly emitted from multiple paths, or \emph{Inconsistent Logging}, where logged data may lack of control.

\subsubsection{Source-Code-Level Validation of Event Parameter Constraints}

Bytecode-level analysis is limited as it only reveals indexed event parameters, without access to the full semantics of events.
Consequently, bytecode-level analysis can only detect potential \emph{Inconsistent Logging} issues by identifying inconsistencies in indexed parameters, and it cannot address non-indexed parameters.
To achieve complete validation, source code analysis is necessary.
At the source code level, we gain access to both indexed and non-indexed parameters, enabling a comprehensive evaluation of potential vulnerabilities, including both \emph{Event Counterfeiting} and \emph{Inconsistent Logging}, which bytecode-level analysis alone cannot fully uncover.

To address these challenges, we rely on the full semantic context of the contract source code to trace event-related call paths and apply symbolic execution to verify parameter constraints.
This ensures that events are emitted with consistent values across all paths, aligning both indexed and non-indexed parameters with the intended contract logic. Additionally, \emph{Event Counterfeiting} detection at the source-code level involves analyzing multiple execution paths to verify that events are not improperly used across functions, providing a depth of validation that cannot be achieved with bytecode analysis alone.

\begin{algorithm}
\footnotesize
  \caption{Finding Inconsistent Logging and Event Counterfeiting via Symbolic Execution}\label{alg:source}
  \begin{algorithmic}[1]
    \Statex \textbf{Input:} \textit{SC}, the source code of smart contracts.
    \Statex \textbf{Output:} $E_v$, a set of phantom events.
      \State $E_v = \emptyset$
      \State \textbf{extract} $E$, i.e., all events from SC. \label{source:event}

      \For{$event \in E$}
        \State $Paths \gets \textsc{SearchPaths}({event, SC})$ \Comment{Get all paths reaching the event}

        \For{$p \in Paths$}
          \State $constraints \gets \textsc{symbolicExec}({p})$
          \If{\Call{CheckLoggingInconsistency}{constraints}}
            \State{$E_v \gets E_v \cup \{event\}$}
          \EndIf
        \EndFor

        \For{$p_1, p_2 \in Paths$}
          \State $constraints_1 \gets \textsc{symbolicExec}({p_1})$
          \State $constraints_2 \gets \textsc{symbolicExec}({p_2})$

          \If{$\textsc{SMT-Solve}({constraints_1 \land constraints_2})$}
            \State{$E_v \gets E_v \cup \{event\}$}
          \EndIf
        \EndFor
      \EndFor

      \State \Return $E_v$

  \end{algorithmic}
\end{algorithm}

At the source-code level, as detailed in~\cref{alg:source}, the tool begins by initializing an empty set \(E_v\) to store potentially vulnerable events. Then extracts all events \(E\) from the smart contract source code \textit{SC}. For each event in \(E\), the tool performs the following steps:

First, the tool extracts all execution paths \(Paths\) in \textit{SC} that lead to the emission of the event using the \textsc{SearchPaths} function. This involves traversing the contract's call graph to capture all inter-procedural paths from user-callable functions to the event-emitting statements. For each path \(p \in Paths\), the tool performs symbolic execution using \textsc{symbolicExec} to extract the path constraints related to the event parameters. The tool then applies the \textsc{CheckLoggingInconsistency} function to verify if any path shows inconsistencies in logging by comparing constraints for all parameters (both indexed and non-indexed). This function is similar to the bytecode analysis, it will detect vulnerabilities by checking for missing constraints on variables, unchecked external calls, or missing storage operations. If inconsistencies are detected, the event is added to \(E_v\) as potentially vulnerable to \emph{Inconsistent Logging}.

Next, the tool performs \emph{Event Counterfeiting} detection by considering all pairs of distinct paths \((p_1, p_2)\) in \(Paths\) which have the same event. For each pair, it performs symbolic execution to obtain constraints \(constraints_1\) and \(constraints_2\) for each path. Using an SMT solver, it checks if the conjunction \(constraints_1 \land constraints_2\) is satisfiable. If the solver finds a solution, this indicates that there is an overlap in parameter values between the two paths, making it difficult for validators to distinguish between events emitted from different paths. In such cases, the event is added to \(E_v\) as potentially susceptible to \emph{Event Counterfeiting}.

Finally, the tool returns the set \(E_v\), which contains events flagged for potential vulnerabilities. For example, consider the \emph{Deposit} event from~\cref{code:1}, which is emitted by the functions \emph{deposit} and \emph{depositETH}. The tool identifies these functions in the AST of the contract, then analyzes their respective execution paths to extract the constraints for the event parameters. For \emph{depositETH}, the constraint is \(msg.value > 0\), while for \emph{deposit}, the constraint is \(\text{bool}(\text{token.call}(\text{signature})) \land (amount > 0)\). Using an SMT solver, the tool evaluates these constraints to check for any overlapping parameter values. If an intersection exists, this suggests potential \emph{Event Counterfeiting}, as a validator may be unable to distinguish between emissions from these two functions, leading the tool to flag the event as forgeable for further review.
\section{Implementation and Experiments}

\begin{figure}[t]
    \centering
    \begin{subfigure}[b]{\linewidth}
        \centering
        \includegraphics[width=\linewidth]{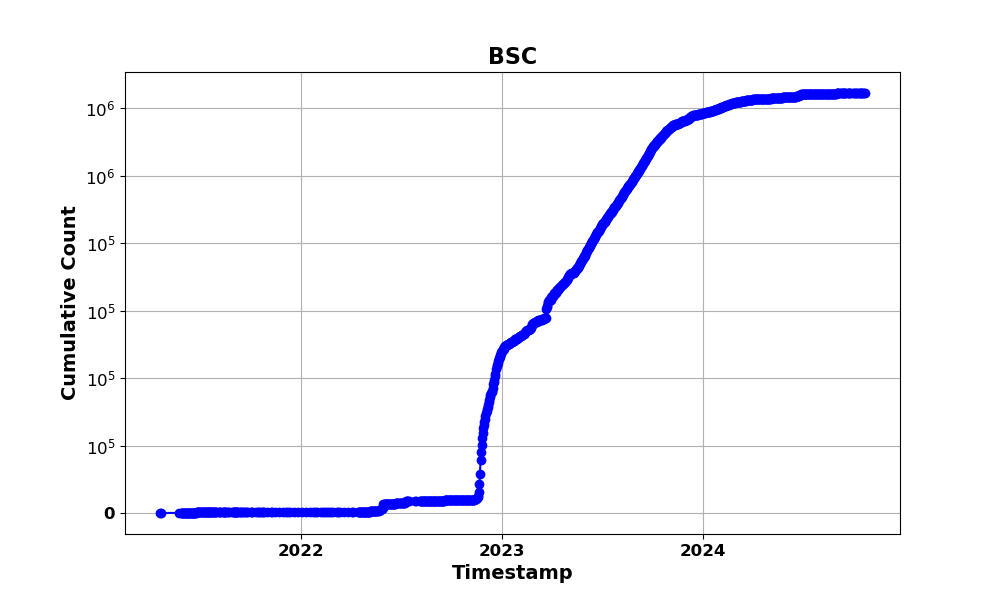}
        \caption{BSC}
        \label{fig:bsc}
    \end{subfigure}
    \vfill
    \begin{subfigure}[b]{\linewidth}
        \centering
        \includegraphics[width=\linewidth]{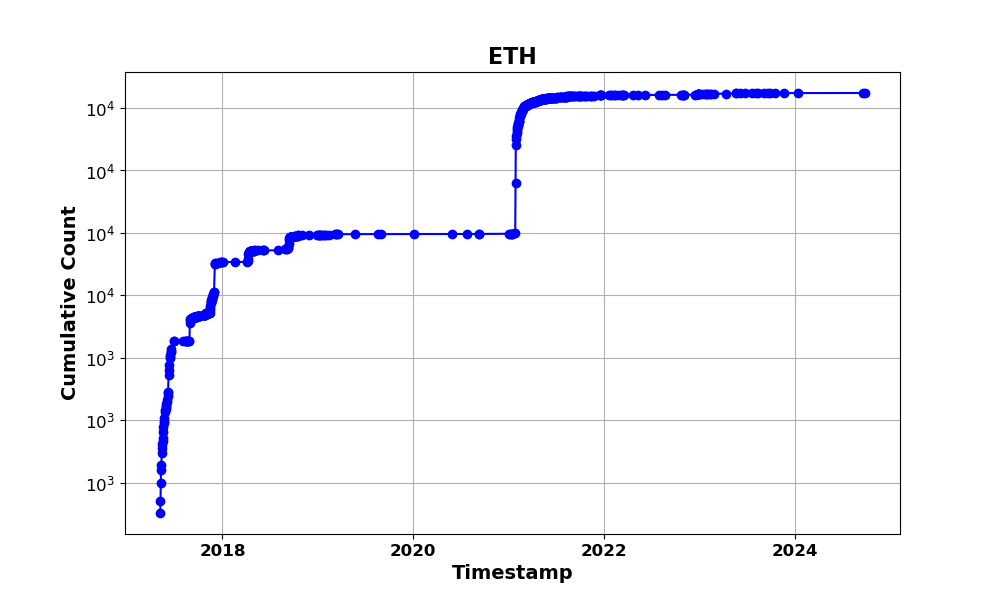}
        \caption{ETH}
        \label{fig:eth}
    \end{subfigure}
    \vfill
    \begin{subfigure}[b]{\linewidth}
        \centering
        \includegraphics[width=\linewidth]{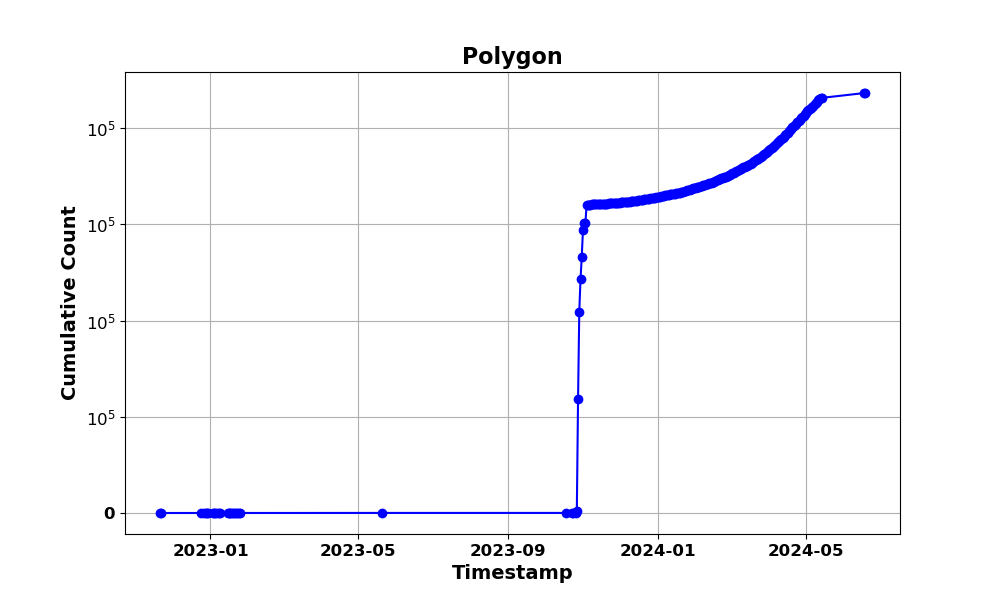}
        \caption{Polygon}
        \label{fig:polygon}
    \end{subfigure}
    \caption{Analysis of ERC-20 transfer phantom events across different platforms for 9 addresses.}
    \label{fig:phantom_events_across_platforms}
\end{figure}

\label{sec:evaluation}
Our tool is implemented in Python and supports both bytecode-level and source code-level analysis for smart contracts, as well as transaction-level monitoring. For bytecode-level analysis, we use the Gigahorse framework to construct the ICFG from compiled bytecode. At the source code level, our tool supports all versions of Solidity compatible with Slither~\cite{Slither}, allowing us to extract the ASTs and perform detailed analysis on event emissions and parameter constraints. We employ the Z3~\cite{Z3} as our constraint solver to handle symbolic execution and eliminate infeasible paths. Additionally, the tool includes a transaction-level off-chain monitoring system that parses on-chain data and applies custom rules to detect anomalies and verify the success of attack transactions. All major components, including the bytecode analysis engine, source code symbolic executor, and transaction monitoring system, were developed by us. The implementation is based in Python and includes over 5,000 lines of code.

\begin{itemize}
  \item \textbf{RQ1} \textit{How prevalent are phantom events on major blockchain platforms, and how does \tool perform in transaction-level detection?} We aim to understand the need for validators to verify the origins of events through contract addresses and source functions.
  \item \textbf{RQ2} \textit{How effective is \tool in detecting potential phantom events from smart contract?} We developed \tool and constructed a custom benchmark to evaluate its effectiveness.
  \item \textbf{RQ3} \textit{How feasible is it to perform such attacks in real world?} We aim to validate whether the identified vulnerabilities can be performed in real-world scenarios.
\end{itemize}

\subsection{Prevalence of Phantom Events (RQ1)}

Tools like TokenScope~\cite{TokenScope} have contributed to detecting inconsistencies in token behavior, and Guoyi Ye et al.~\cite{wallet_visual} have demonstrated the prevalence of forged user addresses for transfer events. We aim to explore two specific aspects: the prevalence of suspicious addresses in phantom event attacks, and whether such attacks are common in cross-chain bridges.

\subsubsection{Methodology}

To better understand the real-world occurrence of phantom events, we conduct two separate experiments using blockchain transaction data.

In the first experiment, we focused on nine particular forged addresses, specifically from \emph{0x1111...1111} through \emph{0x9999...9999}, as these addresses are highly unlikely to be generated or used in normal transactions. We examined the transactions up to October 21, 2024, on the Ethereum, BSC, and Polygon blockchains, looking for ERC-20 token movements originating from these addresses. This approach helps measure the prevalence of phantom events across different blockchain platforms.

In the second experiment, we created detection rules for cross-chain bridges based on our transaction-level analysis, given that bridges are highly vulnerable to phantom event attacks. We compared our findings with those of XScope~\cite{Xscope}, the only existing tool for detecting bridge attack transactions, using its associated dataset.

\subsubsection{Result}


In our first experiment, focusing on nine specific addresses, we identified a significant occurrence of phantom events in ERC-20 token transfers, as shown in \cref{fig:phantom_events_across_platforms}. We found 18,099 such events on Ethereum, 1,245,125 on BSC, and 436,407 on Polygon. Such events, if not properly verified, could lead to security exploits. This significant presence from just nine addresses highlights the critical need for rigorous transaction verification to maintain the security and trustworthiness of blockchain operations.


\begin{table}[t]
    \centering
    \caption{Number of detected attack transactions and successful attacks by tools.}
    \label{tab:attack_detection}
    \begin{tabular}{l r r r}
        \toprule
        \textbf{Project} & \textbf{XScope} & \multicolumn{2}{c}{\textbf{\tool}} \\
        \cmidrule(lr){2-2} \cmidrule(lr){3-4}
        & \textbf{Attacks} & \textbf{Attack Trans.} & \textbf{Successful Attacks} \\
        \midrule
        THORChain \#1 & 9 & 9 & 9 \\
        THORChain \#2 & 41 & 48 & 41 \\
        pNetwork & 3 & 5 & 3 \\
        Qubit Bridge & 20 & 20 & 20 \\
        meter.io & N/A & 5 & 5 \\
        CENNZnet & N/A & 1 & 1 \\
        \bottomrule
    \end{tabular}
\end{table}

In the second experiment (the result shown in \cref{tab:attack_detection}), we detected all transactions reported by XScope and identified additional attack transactions. XScope only records transactions where attacks have already taken place, indicated by a fund transfer on the destination chain. In contrast, we were able to detect attacks before a transfer occurred, helping project teams identify attackers proactively. We also detected attack transactions on meter.io~\cite{Meter}, which XScope mentioned in its attack incident list but did not analyze. Additionally, we discovered an attack transaction on CENNZnet, which resulted in a loss of 150 ETH. This incident had not been previously disclosed or reported by any other tools or security companies.

\vspace{\baselineskip}
\begin{mdframed}
\noindent\textbf{Answer to RQ1:} \textit{Our analysis reveals a significant presence of phantom events across Ethereum, BSC, and Polygon, based on just nine addresses. Furthermore, our detection rules demonstrated effectiveness compared to state-of-the-art tools.}
\end{mdframed}

\subsection{Phantom Event Detection from Code (RQ2)}

\subsubsection{Datasets}
SmartAxe~\cite{liao2024smartaxe} and XGuard~\cite{wang2024xguard} are currently the two known tools for detecting cross-chain vulnerabilities, with SmartAxe focusing on bytecode-level analysis and XGuard providing source code-level analysis.

In our analysis of the SmartAxe dataset, we identified that only 8 of the 20 attacks were caused by \emph{Event Counterfeiting}, with vulnerable function path pairs existing within bridge contracts. We re-labeled these 8 attacks and added 29 more function path pairs affected by \emph{Event Counterfeiting} from the 129 bridges analyzed by SmartAxe. To complement this, we selected 500 BSC smart contracts with low similarity (using the TF-IDF algorithm) from the 452,666 verified contracts in BSC~\cite{bscdata} and manually annotated them. After deduplication, we identified 80 additional function pairs affected by \emph{Event Counterfeiting}.

By combining SmartAxe and XGuard datasets, along with manually annotated function pairs, we assembled a comprehensive data set containing 117 function pairs affected by \emph{Event Counterfeiting} for evaluating our detection tool.

To detect \emph{Inconsistent Logging}, we used a dataset of 28 manually audited cases of \emph{Inconsistent Logging} parameters, which were identified during our audits.

\subsubsection{Result}

\begin{table*}[htbp]
    \centering
    \caption{Tools Comparison for Event Counterfeiting and Inconsistent Logging}
    \label{tab:tool_result}
    \small
    \begin{tabular}{p{3.9cm}ccc ccc}
        \toprule
        \multirow{2}{*}{\textbf{Tool}} & \multicolumn{3}{c}{\textbf{Event Counterfeiting}} & \multicolumn{3}{c}{\textbf{Inconsistent Logging}} \\
        \cmidrule(lr){2-4} \cmidrule(lr){5-7}
        & \textbf{Precision} & \textbf{Recall} & \textbf{F1} & \textbf{Precision} & \textbf{Recall} & \textbf{F1} \\
        \midrule
        SmartAxe & 40.50\% & 25.12\% & 31.01\% & N/A & N/A & N/A \\
        XGuard & 87.75\% & 55.12\% & 67.70\% & 64.71\% & 91.75\% & 75.82\% \\
        \tool (Bytecode) & 84.0\% & 100\% & 91.30\% & 62.50\% & 16.12\% & 39.31\% \\
        \tool (Source) & 90.83\% & 86.51\% & 88.62\% & 90.30\% & 100\% & 94.90\% \\
        \bottomrule
    \end{tabular}
\end{table*}
In our experiments, \tool significantly outperformed SmartAxe and XGuard in detecting both \emph{Event Counterfeiting} and \emph{Inconsistent Logging} vulnerabilities, as shown in \cref{tab:tool_result}. For \emph{Event Counterfeiting}, SmartAxe achieved an F1 score of 31.01\% with a precision of 40.50\% and a recall of 25.12\%, while XGuard reached an F1 score of 67.70\% with a precision of 87.75\% and a recall of 55.12\%. In comparison, \tool demonstrated superior performance, achieving an F1 score of 91.30\% at the bytecode level and 88.62\% at the source code level. For \emph{Inconsistent Logging} detection, \tool also outperformed XGuard, achieving an F1 score of 94.9\% with perfect recall, compared to XGuard's 75.82\%.

The difference in performance metrics for \tool’s bytecode and source code implementations highlights specific limitations in each approach. In \emph{Event Counterfeiting} detection, \tool’s bytecode-level analysis lacks constraint validation, leading to a lower precision as it may flag events that lack sufficient evidence of forgery. However, at the source code level, \tool performs comprehensive constraint checks, improving precision significantly. For \emph{Inconsistent Logging}, \tool’s bytecode analysis is limited to indexed parameters only, which reduces recall as it cannot identify issues related to non-indexed parameters. With source code access, \tool can analyze both indexed and non-indexed parameters, thus achieving higher recall. In contrast, SmartAxe and XGuard have inherent design limitations; SmartAxe only detects whether two functions emit the same event without considering constraint validation within call paths, while XGuard focuses on constraints within single functions and lacks inter-function analysis. By analyzing both function interactions and call paths, \tool detects vulnerabilities that the other tools miss, achieving superior results in both \emph{Event Counterfeiting} and \emph{Inconsistent Logging} detection.

\subsubsection{Limitations}

The false positives and false negatives in our evaluation revealed two main limitations of our approach. First, when dealing with incomplete contract code, such as interface calls in certain functions, our computed constraints may not always be precise, leading to weaker constraints and thus false positives. Second, compilation errors occurred with some smart contracts that were verified by blockchain explorers but failed to compile with our tool for analysis~\cite{compile_issue}, due to Solidity compile bugs. Addressing these limitations will be a focus of our future work.

\vspace{\baselineskip}
\begin{mdframed}
\noindent\textbf{Answer to RQ2:} \textit{\tool has proven highly effective in detecting phantom event vulnerabilities, surpassing current state-of-the-art tools in precision, recall, and F1 score.}
\end{mdframed}

\subsection{Real-World Attack Feasibility (RQ3)}
\subsubsection{Methodology}
We applied our tool to conduct testing on various blockchain platforms, aiming to identify vulnerabilities related to phantom event attacks in smart contracts. Beyond on-chain testing, we also audited various off-chain applications, including blockchain explorers, NFT marketplaces, and cryptocurrency wallets. These audits focused on evaluating security protocols and identifying potential vulnerabilities that could be exploited by phantom events, ensuring a comprehensive analysis of both on-chain and off-chain components.

All identified phantom event vulnerabilities were tested in a controlled local environment using simulated scenarios to ensure no disruption or harm to real-world blockchain platforms.

\subsubsection{Result}

In our experiments, we identified a \emph{Transfer Event Spoofing} transaction in which a fake transfer appeared to move tokens from the address \textit{0x8888...8888} to the victim's wallet. Despite no actual token transfer occurring, at least five major blockchain explorers—BscScan, OKLink, Bitquery, Tokenview, and Bsctrace~\cite{bscscan,oklink,bitquery,tokenview,bsctrace}—incorrectly recognized this spoofed event as a legitimate transaction. This highlights the widespread vulnerability in off-chain systems, which struggle to differentiate between genuine and phantom events.

Additionally, we successfully replicated the ``sleepminting'' attack on testnet environments for two leading NFT marketplaces, Opensea and Rarible. Despite the existence of monitoring solutions designed to detect such attacks, fully mitigating these vulnerabilities remains an ongoing challenge.

Moreover, during our security audits, we discovered critical vulnerabilities in six cryptocurrency wallets, which we reported to the respective project teams or through bug bounty platforms. Four of these vulnerabilities were confirmed, with one resulting in a \$600 bounty from the project team.

In our audit of several blockchain bridges, we identified vulnerabilities in off-chain code that left them exposed to the Mimicry Contract Attack. In this attack, malicious contracts emit forged events that are incorrectly processed as legitimate by off-chain systems. One vulnerable system we uncovered was forked and deployed by a public blockchain project with a market capitalization exceeding \$250 million as of October 9, 2024. These vulnerabilities have been reported to the respective project teams for remediation.

Additionally, we identified a display issue with event data in the popular blockchain explorer Blockscout~\cite{error_display}, which could compromise the accuracy of transaction records. This flaw could potentially lead to user misinterpretation of blockchain data, undermining trust in transaction history.

In our audit of multiple active smart contract projects, we also found that three DeFi projects and one GameFi project were vulnerable to the \emph{Inconsistent Logging} attack, a vulnerability first proposed in this paper (see \ref{vec:attack2}). The highest affected market capitalization in this set of projects was \$169,688.

\vspace{\baselineskip}
\begin{mdframed}
\noindent\textbf{Answer to RQ3:} \textit{Our research demonstrates that phantom event attacks are feasible on real-world blockchain platforms, specifically targeting blockchain explorers, NFT marketplaces, DeFi, GameFi, and cryptocurrency wallets. This emphasizes the urgent need for improved security measures within the blockchain ecosystem. Detailed reports and confirmation cases are available in the supplementary materials.}
\end{mdframed}

\subsection{Threats to Validity}
The internal validity threats primarily stem from potential inaccuracies introduced during the manual data labeling process. To improve the accuracy of data labeling, we adopted a three-tier approach, where two authors independently labeled the data and a third author reviewed their labels. This method involved a thorough review at three distinct levels to enhance the precision of our dataset categorization.

To ensure external validity, we diversified the types and sources of datasets used in our experiments. In addition to the bridge-focused data provided by SmartAxe and XGuard, we expanded our dataset to include a broader range of scenarios. Furthermore, we avoided using smart contracts with high similarity to improve the diversity and validity of our analysis.
\section{Related Work}
\subsection{Smart Contract Security Analysis}
In recent years, a large number of techniques have been proposed to analyze the security of smart contracts.

\paragraph{Static Analysis}
Tikhomirov et al.~\cite{tikhomirov2018smartcheck} designed SmartCheck, a system translating
Solidity source code to XML and detecting bugs via xPath patterns. Grech et
al.~\cite{grech2022elipmoc} proposed a static analysis framework termed Gigahorse that translates
the stack-based bytecode to the register-based intermediate representation. A similar tool named
Slither~\cite{Slither} was proposed by Fesit that targets solidity source code. Lu et
al.~\cite{lu2021neucheck} design a smart contracts security analysis tool named NeuCheck which is
based on syntax tree parsing. Ghaleb et al.~\cite{ghaleb2022etainter} proposed eTainter, a static analysis tool that detects gas-related vulnerabilities in smart contracts by applying taint tracking to bytecode.

\paragraph{Symbolic Execution}
Luu et al.~\cite{luu2016making} proposed Oyente, a pioneering symbolic execution tool for Ethereum smart contracts to detect bugs. Lin et al.~\cite{lin2022solsee} presented SolSEE, the first source-level symbolic execution engine for Solidity smart contracts. Ma et al.~\cite{ma2021pluto} introduced Pluto for detecting security bugs by reconstructing inter-contract CFGs. Pasqua et al.~\cite{pasqua2023enhancing} proposed a method based on the symbolic execution of EVM operands for precise CFG construction and improved vulnerability detection. Ruaro et al.~\cite{ruaro2024not} implement CRUSH, which leverages symbolic execution and program slicing to detect storage collisions among such contract groups. Gritti et al.~\cite{gritti2023confusum} developed JACKAL, which performs symbolic execution based on control flow graphs (CFG) and function call graphs (FCG) to detect confused contract vulnerability. In addition, industry solutions such as Mytrhil~\cite{mythril} and Manticore~\cite{mossberg2019manticore} have become standard tools for smart contract audits.

Our approach involves developing a detection framework that combines multiple techniques, including bytecode-level analysis, source code analysis, transaction-level monitoring, and symbolic execution, but the issues addressed in this paper cannot be captured by any existing vulnerability patterns.

\subsection{Smart Contract Events}
Generally, logging messages enhances program comprehension and reduces maintenance costs, but
research on Solidity event logging and security is still limited. Li et
al.~\cite{li2023understanding} conducted the first empirical study on Solidity event logging and
developing a tool to identify the event that causes unnecessary gas usage. Zhang et
al.~\cite{Xscope} designed a tool named Xscope which finds
security violation events in cross-chain bridges. Cernera et al.~\cite{cernera2023token} tracked
the tokens created by internal transactions by scanning the logs looking for Transfer events.
Additionally, Guidi et al.~\cite{sleepminting} analyzed the phenomenon of sleepminting and explored the use of Forta for
tracking and alerting about suspicious events.
Zhu et al.~\cite{doccon} proposed a technique called DocCon, which detects inconsistencies between
Solidity code and its corresponding documentations. These inconsistencies include documented event
emissions which do not appear in code, or vice versa.
TokenScope \cite{TokenScope} contributed to the detection of inconsistencies in token behaviors by monitoring ERC-20 events.
Despite these advancements, most research has only considered issues related to events in specific
scenarios and has not fully addressed the generality of attacks caused by phantom events.

\label{sec:related}
\section{Mitigation Strategies}
\label{sec:mitigation}

Mitigating vulnerabilities caused by phantom events requires a comprehensive approach, addressing smart contract development, ecosystem infrastructure, and attack detection mechanisms. From the perspective of contract development, developers should implement strict validation mechanisms to ensure that event parameters are verified before emission and access control mechanisms for the functions. It is essential to enforce proper state transitions to prevent mismatches between emitted events and the actual contract state. 

At the ecosystem level, off-chain systems like blockchain explorers, wallets, and DApps must adopt more robust validation techniques to distinguish legitimate events from phantom events. Event emitter validation, where the source of the event is cross-checked with the contract address, helps ensure that events originate from authorized contracts. Furthermore, improving data sanitization processes in off-chain applications is critical to prevent vulnerabilities such as cross-site scripting (XSS) and SQL injection (SQLi). Enhanced cross-chain security protocols are necessary for cross-chain bridges, ensuring that events on both the source and destination chains are validated to prevent event forgery and manipulation.

In terms of security attack detection, continuous real-time monitoring of on-chain transactions and events is essential to detect and flag suspicious activities, such as \emph{Transfer Event Spoofing} or \emph{Contract Imitation}. Defining detailed detection rules, both for on-chain contract behavior and off-chain event handling, allows for more comprehensive identification of vulnerabilities. Additionally, regular security audits of both smart contracts and off-chain systems should be conducted to identify potential weaknesses, particularly focusing on event emission logic, access controls, and transaction validation. Through a combination of these strategies, the risk posed by phantom events can be significantly reduced, improving the security and reliability of blockchain systems.

\section{Conclusion}

In this work, we conducted an in-depth analysis of vulnerabilities associated with phantom events in blockchain systems. Our approach involved developing a multi-level detection framework that integrates bytecode-level, source code-level, and transaction-level analyses to address issues such as \emph{Event Counterfeiting}, \emph{Inconsistent Logging}, and \emph{Contract Imitation}. Furthermore, we identified real-world cases for each of the proposed attack vectors, underscoring the critical need for mitigation of these threats in blockchain ecosystems.



\appendices


\section{Real-World Attack Examples}

\subsection{Examples of the Five Attack Vectors in Our Taxonomy}
\label{appendix:examples}

\paragraph{Event Counterfeiting}
Two well-known instances of \emph{Event Counterfeiting} are the attacks on Qubit Bridge~\cite{Qubit} and Meter Bridge~\cite{Meter}, which resulted in losses of \$80 million and \$4.3 million, respectively.

\paragraph{Inconsistent Logging}
Many of these issues were discovered from the 25 most active smart contracts on the BSC chain on 1 April 2024. Specifically, the contracts for TroyEmpire~\cite{TroyEmpire}, SecondLive~\cite{SecondLive}, WalletLocking~\cite{WalletLocking}, and TTGameEvents~\cite{TTGame} were found to have these issues. For example, TTGame project generates an address for users' registration and transfers a certain amount of BNB to the address as gas fees for future automatic calls. By identifying authorized user addresses from the historical transactions, we can find the remaining unauthorized transactions that emit phantom events. Although user permissions for the other three projects were unclear, we can simulate to call their functions and arbitrarily generate withdrawal events, confirming the existence of these issues.

\paragraph{Contract Imitation}
For \emph{Contract Imitation} attack, the most notable instance is the attack on pNetwork, which resulted in a loss of \$12.7 million. This has already been briefly introduced in \cref{sec:motivation}. Similarly, THORChain experienced a loss of \$8 million due to an error handler issue~\cite{thorchain}. As for mimicry contract attack, a representative case is the ``sleepminting'' attack, which gained attention for generating an NFT valued at \$69 million and listing it on a platform. This incident has been widely studied in the academic community~\cite{nft_sleepminting,clone_nft,sleepminting}. According to recent information from the Forta platform~\cite{sleepminting_forta}, there were 8,130 sleepminting alerts generated in just 7 days, from December 3 to December 9, 2023.

\paragraph{Transfer Event Spoofing}
Our analysis of on-chain data revealed numerous instances of \emph{Transfer Event Spoofing}, where ERC-20 token and NFT events had altered sender addresses. These included forged addresses resembling celebrities, well-known exchanges, and other eye-catching entities. Several of these cases have led to significant financial losses~\cite{layerzero_scam,TetherClaims,ethaddress,zero_transfer,wallet_visual}.

\paragraph{Event Handling Error}
A well-known example of \emph{Event Handling Error} involves the misinformation case experienced by Rarible and OpenSea, caused by sleepminting. Blockchain explorers like Etherscan and Bscscan, as well as various wallets, often treat forged transfer events as legitimate transactions. Another instance involves blockwell.ai, where wallets mistakenly identified a phantom event from a spoofed token as a valid ERC-20 transfer~\cite{blockwell.ai}. Similar vulnerabilities have been observed in the ERC-721 and ERC-1155 tokens.

In addition, OpenSea, the largest NFT marketplace, has faced similar issues. The attackers manipulated the metadata of the NFTs displayed by OpenSea, embedding malicious payloads that triggered unwanted wallet behaviors, ultimately allowing attackers to profit. Similar attacks have been found in ERC-20 projects, highlighting the widespread difficulty in securely handling event data across the blockchain ecosystem~\cite{rektosaurus,OpenSea}.

\subsection{Framework of Inconsistent Logging and Contract Imitation}

\begin{figure}
    \centering
    \includegraphics[width=1\linewidth]{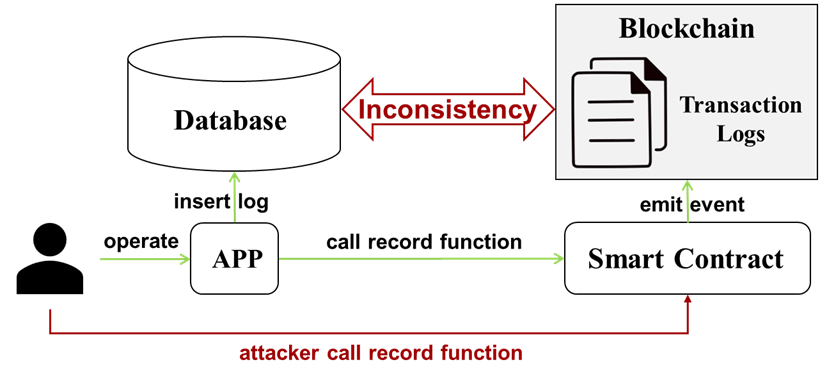}
    \caption{The overview of Inconsistent logging apps.}
    \label{fig:Inconsistent_logging}
\end{figure}

~\Cref{fig:Inconsistent_logging} provides an overview of \emph{Inconsistent Logging} in applications. In this framework, a user operates through an application, which interacts with a database to insert logs and with a smart contract on the blockchain to emit events. However, inconsistencies can arise between the database logs and the blockchain transaction logs due to discrepancies in logging practices. Attackers can exploit this by directly calling the record function, potentially creating mismatches between the application's internal log records and the blockchain's transaction logs.

\begin{figure}[!htbp]
    \centering
    \begin{subfigure}[b]{0.5\textwidth}
        \centering
        \includegraphics[width=\textwidth]{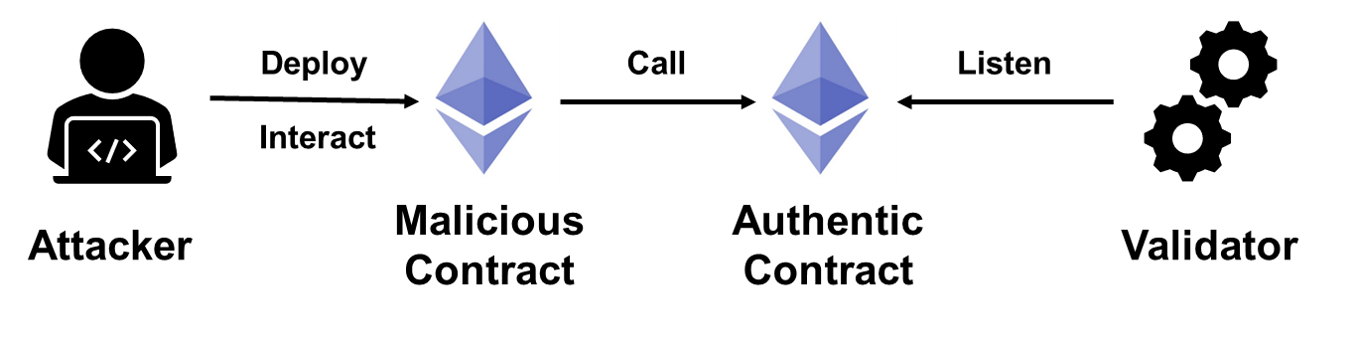}
        \caption{Blended Event Attack}
        \label{fig:attack3-1}
    \end{subfigure}
    \quad
    \begin{subfigure}[b]{0.5\textwidth}
        \centering
        \includegraphics[width=\textwidth]{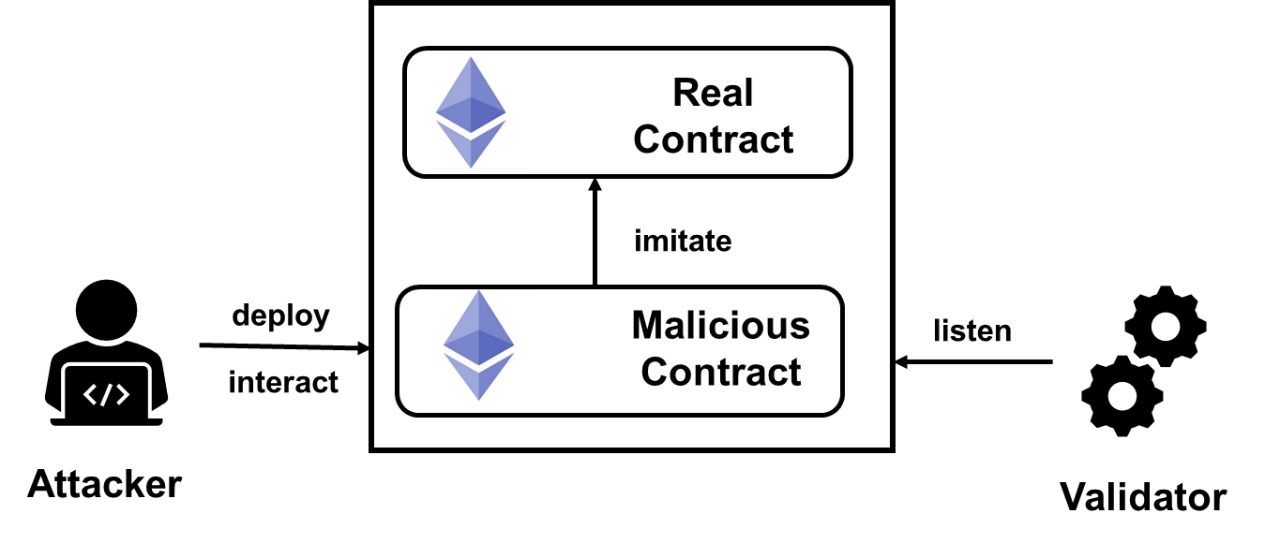}
        \caption{Mimicry Contract Attack}
        \label{fig:attack3-2}
    \end{subfigure}
    \caption{Two type of Contract imitation}
    \label{fig:attack1}
\end{figure}

~\Cref{fig:attack1} illustrates two types of contract imitation attacks. In the \emph{Blended Event Attack}, a malicious contract interacts with an authentic contract, and logs from both contracts are recorded in the same transaction, which may mislead validators. In the \emph{Mimicry Contract Attack}, the attacker deploys a contract that imitates a real contract, generating deceptive logs that appear to originate from the authentic contract, creating further confusion for validators.

\subsection{Case of Transfer Event Spoofing}

\begin{figure}
    \centering
    \includegraphics[width=1\linewidth]{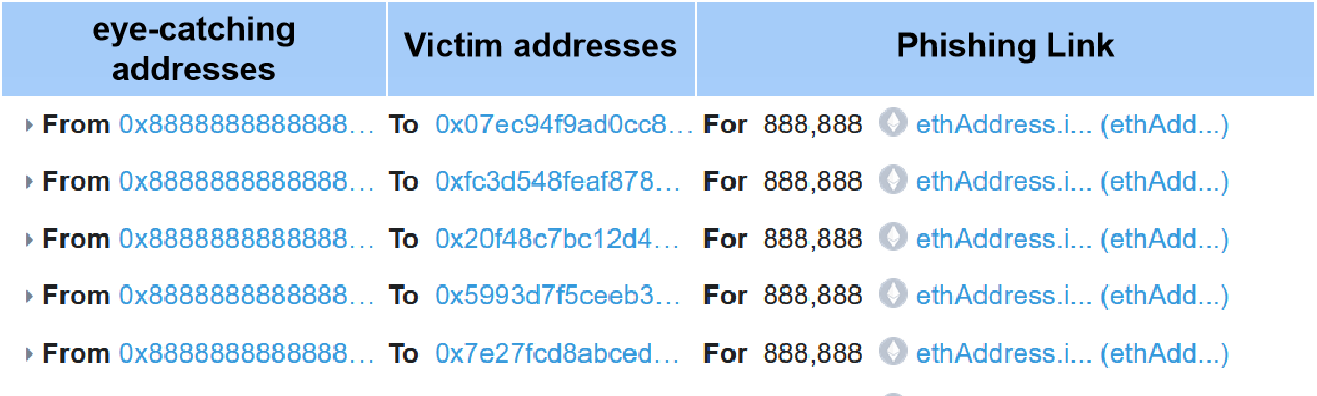}
    \caption{Airdrop event spoofing.}
    \label{fig:airdrop}
\end{figure}

~\Cref{fig:airdrop} illustrates an example of a \emph{Transfer Event Spoofing} attack, where the attacker forges the event sender, causing blockchain explorers and wallets to incorrectly display it as a successful transfer. This misrepresentation can deceive users and is used as a social engineering tactic to exploit trust in these interfaces.

\begin{table}[]
\caption{Real case of transfer event spoofing attack transactions.}
\label{tab:transfer_spoof_attack_transaction}
\begin{tabular}{lllll}
\toprule
  & Sender & Transfer From                & Transfer To                  & Token      \\
\midrule
1 & Victim           & Victim & 0x{\color[HTML]{FE0000}734}659...50C{\color[HTML]{FE0000}a79F7} & USDT \\
\midrule
2 & Attacker              & Victim & 0x{\color[HTML]{FE0000}734}35A..2Bc{\color[HTML]{FE0000}a79F7}  & FAKE USDT     \\
\midrule
3 & Victim           & Victim & 0x{\color[HTML]{FE0000}734}35A...2Bc{\color[HTML]{FE0000}a79F7} & USDT \\
\bottomrule
\end{tabular}
\end{table}
The \emph{Transfer Spoofing} attack on 9 February 2024, where an attacker successfully executed a \emph{Transfer Event Spoofing} attack, resulting in a loss of \$1.04 million USD. This attack exploited vulnerabilities in wallets and blockchain explorers that parse and display transfer events. As shown in~\Cref{tab:transfer_spoof_attack_transaction}, after a user sent funds to a legitimate address (0x734\ldots{}a79F7), the attacker forged a transfer event, making it appear as though the funds were sent to a similar address controlled by the attacker. Due to the way certain wallets and explorers processed these events, the fake transfer was displayed as legitimate, leading the user to mistakenly send additional funds to the attacker's address, resulting in significant financial losses~\cite{zero_transfer}.

\end{document}